\pdfoutput=1 
\documentclass{JINST}
\usepackage{xspace}
\def\github {\href{http://github.com/jmhardin/FastDIRC}{{\em github.com/jmhardin/FastDIRC}}}

\def\geant    {\mbox{\textsc{Geant}}\xspace}
\def\babar    {\mbox{\textsc{BaBar}}\xspace}
\def\thetac   {\mbox{$\theta_C$}\xspace}

\title{FastDIRC: a fast Monte Carlo and reconstruction algorithm for DIRC detectors}

\author{John Hardin and Mike Williams\\
Laboratory for Nuclear Science, Massachusetts Institute of Technology, Cambridge, MA 02139}

\abstract{
FastDIRC is a novel fast Monte Carlo and reconstruction algorithm for DIRC detectors.
A DIRC employs rectangular fused-silica bars both as Cherenkov radiators and as light guides.
Cherenkov-photon imaging and time-of-propagation information are utilized by a DIRC to identify charged particles.
\geant-based DIRC Monte Carlo simulations are extremely CPU intensive. 
The FastDIRC algorithm permits fully simulating a DIRC detector more than $10\,000$ times faster than using \geant.  This facilitates designing a DIRC-reconstruction algorithm that improves the Cherenkov-angle resolution of a DIRC detector by $\approx 30\%$ compared to existing algorithms. 
FastDIRC also greatly reduces the time required to study competing DIRC-detector designs.
}

\begin{document}

\section{Introduction}

\begin{figure}[]
	\begin{center}
	\includegraphics[height=8.cm]{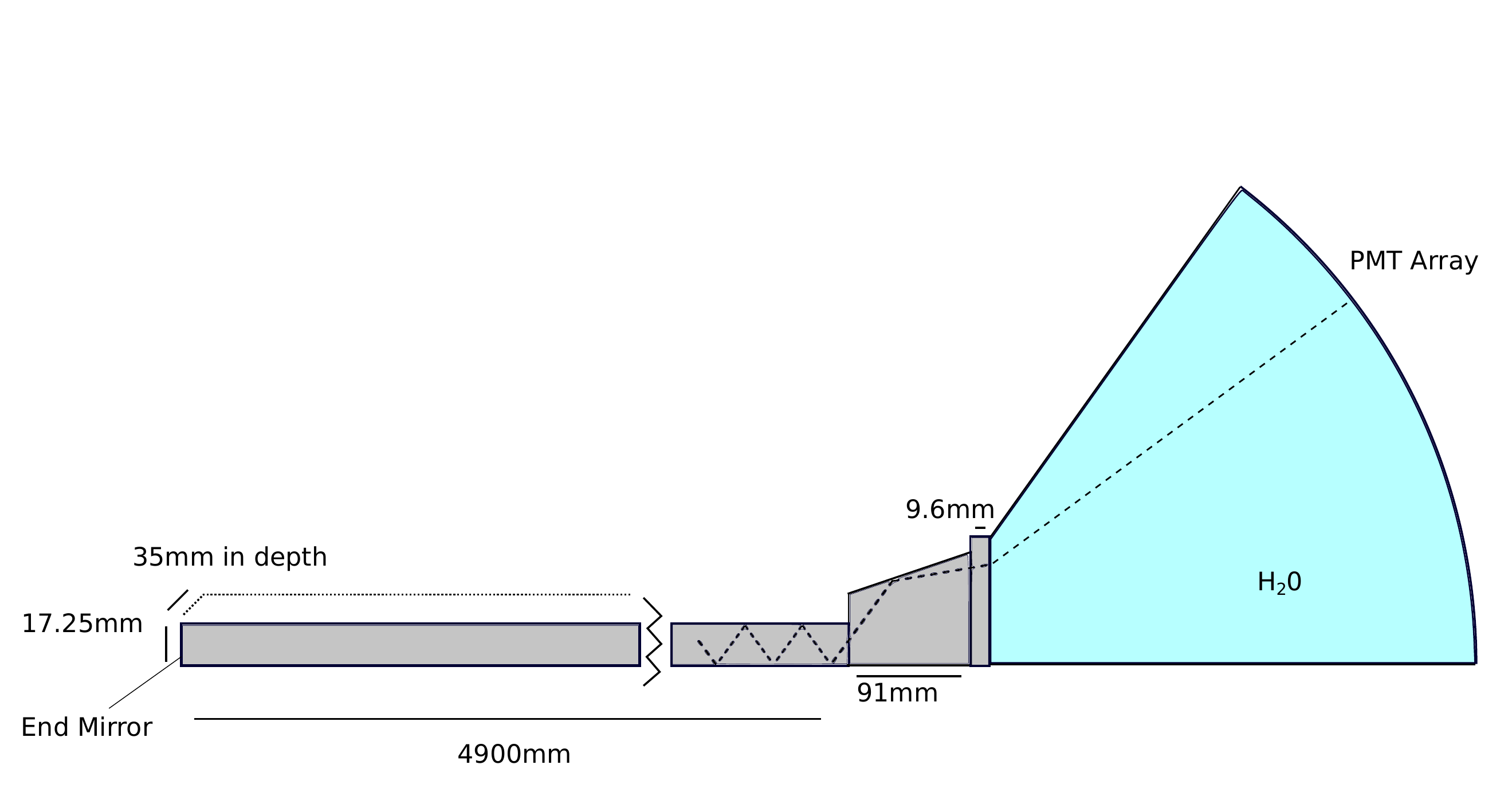}
	\caption{Schematic diagram of one \babar box~\cite{babar} that contains 12 fused-silica bars (each 17.25\,mm thick, 35\,mm wide, and 4.9\,m long) and wedges. \babar used a large tank filled with purified water to transport photons from the fused-silica wedges to the PMT array. 
        Drawing not to scale. 
} 
	\label{fig:babar}
	\end{center}
\end{figure}

The world's first Detection of Internally Reflected Cherenkov light (DIRC) detector was developed and utilized by the \babar experiment at SLAC~\cite{babar}.  
The radiator of the \babar DIRC consisted of a barrel made up of twelve boxes each containing twelve synthetic fused-silica bars. 
The bars also served as light guides for the Cherenkov light trapped inside them by total internal reflection.  
One end of each box was coupled to a photon camera, while the other end has a mirror that reflected the light back to the photon-detector side (see Fig.~\ref{fig:babar}).  
Using the image reconstructed by the photon camera and the three-momentum information provided by the {\sc BaBar} tracking system, the angle at which the Cherenkov light was emitted \thetac was determined with a precision of 2.5\,mrad (9.6\,mrad per photon).  
The \babar DIRC provided excellent particle identification performance up to a momentum of about 4\,GeV$/c$.  

The success of the \babar DIRC has inspired other experiments to consider similar DIRC-like detectors\,\cite{superb,belle2,gluex,torch,panda}.
Various photon-camera designs are being studied, but even the most complicated optics employed in this DIRC component is straightforward to simulate since the Cherenkov photons are only reflected at most a few times (a necessity given current mirror technology).
Simulating the path of a Cherenkov photon that internally reflects $\mathcal{O}(100)$ times within a rectangular bar, however, is extremely CPU intensive. 
This causes two important difficulties: 
(1) optimizing the design of a DIRC detector is challenging due to the CPU required to consider different design choices; and
(2) thus far only approximate pattern-recognition algorithms have been employed, since the CPU cost of generating a unique pattern for each track is too prohibitive.

In this article, we present a novel fast Monte Carlo algorithm (FastDIRC) for simulating DIRC detectors.  
This algorithm is more than $10\,000$ times faster than a \geant-based simulation~\cite{geant,superbmc}, which
permits studying DIRC-detector designs systematically, simplifying their optimization.
Furthermore, FastDIRC facilitates generating a unique probability density function (PDF) -- in both spatial position on the photo-detector (PMT) plane and in time -- for each particle whose Cherenkov angle is to be reconstructed.
We will show that using such a per-track PDF improves the resolution by $\approx 30\%$ compared to the approximate PDFs employed by existing algorithms. 
The outline of this paper is as follows:
existing Monte Carlo and reconstruction methods are discussed in Sec.~2;
FastDIRC is presented in Sec.~3;
Sec.~4 discusses potential improvements;
and we summarize in Sec.~5.

\section{Established Methods}

As a concrete example for discussing both established methods and FastDIRC, we will use the SuperB focusing DIRC (FDIRC) prototype that was designed and tested at SLAC~\cite{superb}. 
This choice is motivated by the wealth of published SuperB FDIRC performance data with which we can compare and validate our simulations.
We stress, however, that FastDIRC is a general algorithm that can be used to simulate any DIRC-like detector.   

\begin{figure}[]
	\begin{center}
	\includegraphics[height=8.cm]{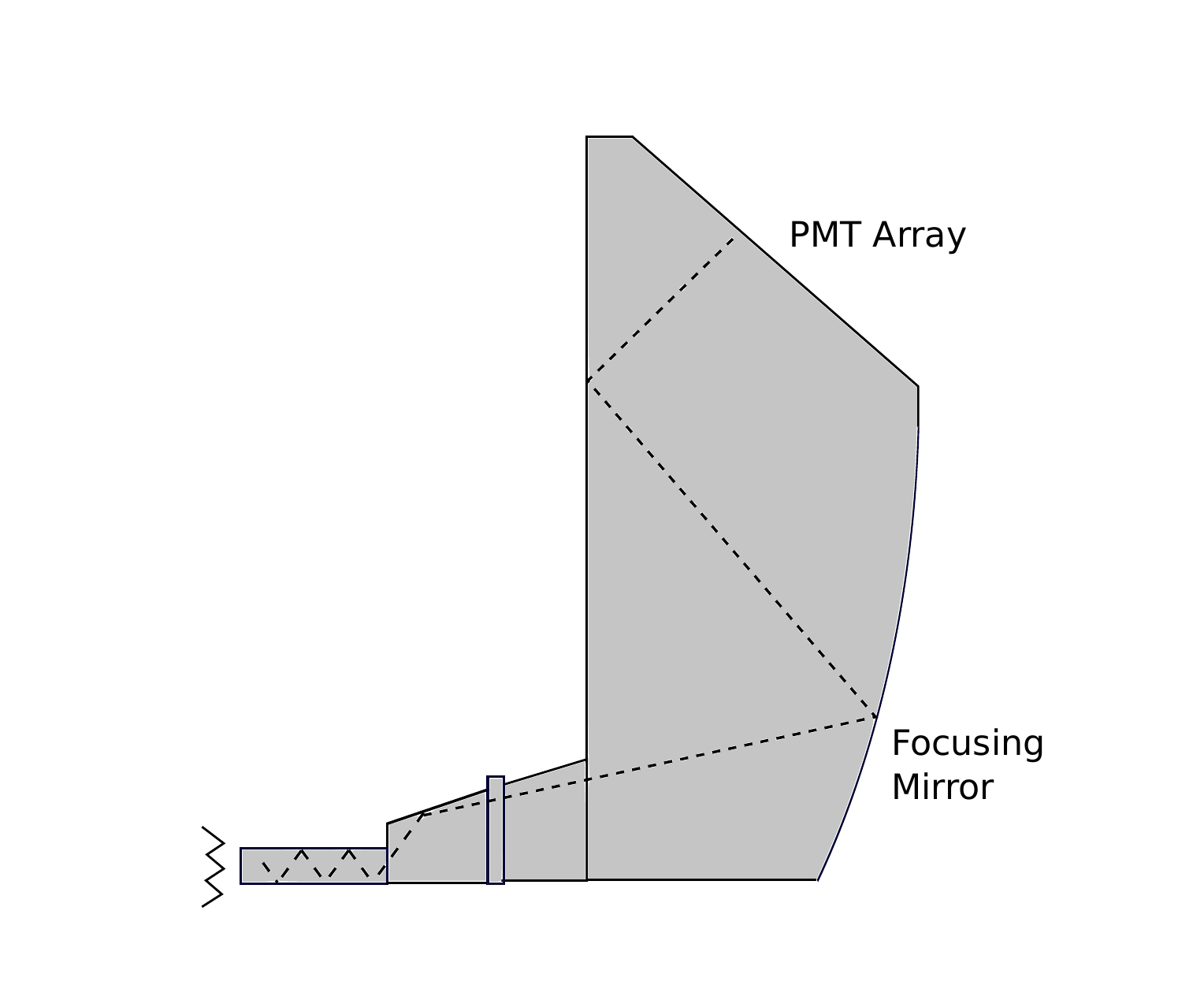}
	\caption{Schematic diagram of the SuperB FDIRC focusing block~\cite{superb} which is attached to a prototype \babar DIRC fused-silica bar box. 
} 
	\label{fig:superb}
	\end{center}
\end{figure}

The SuperB FDIRC prototype utilized a prototype \babar DIRC fused-silica bar box to minimize cost~\cite{superb}. 
A major revolution in the design -- compared to that of the \babar DIRC -- was the use of a fused-silica focusing block instead of a large purified-water-filled expansion volume.
This design greatly reduced the area of PMTs required (hence the cost of the system), and the focusing cylindrical mirror removed the uncertainty due to where the photon was emitted in the bar since parallel rays are focused to the same point on the PMT plane in the focusing direction.
Figure~\ref{fig:superb} shows a schematic of the SuperB FDIRC detector. 

In this section, we will simulate the SuperB FDIRC and reconstruct its Cherenkov light using established methods. 
We do not consider uncertainties that originate in the tracking system; {\em i.e.}, we assume the momentum $\vec{p}$ of each particle is perfectly measured by the tracking system.
It is trivial to include such uncertainties; however, for our purposes here, they only serve to obscure the performance of the DIRC detector and its simulation.

\subsection{\geant Monte Carlo}

\begin{figure}[]
	\begin{center}
	\includegraphics[width=0.49\textwidth]{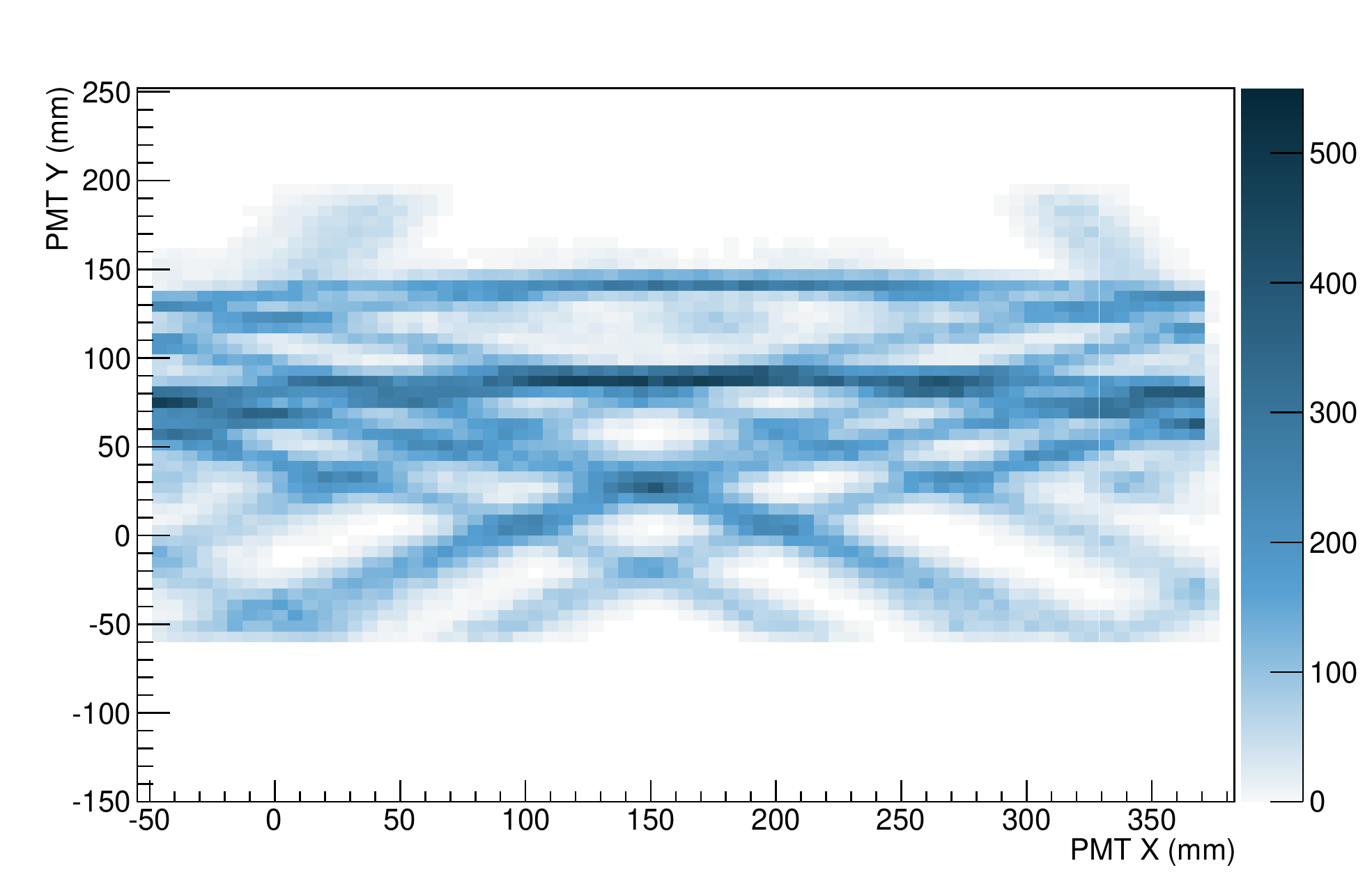}
	\includegraphics[width=0.49\textwidth]{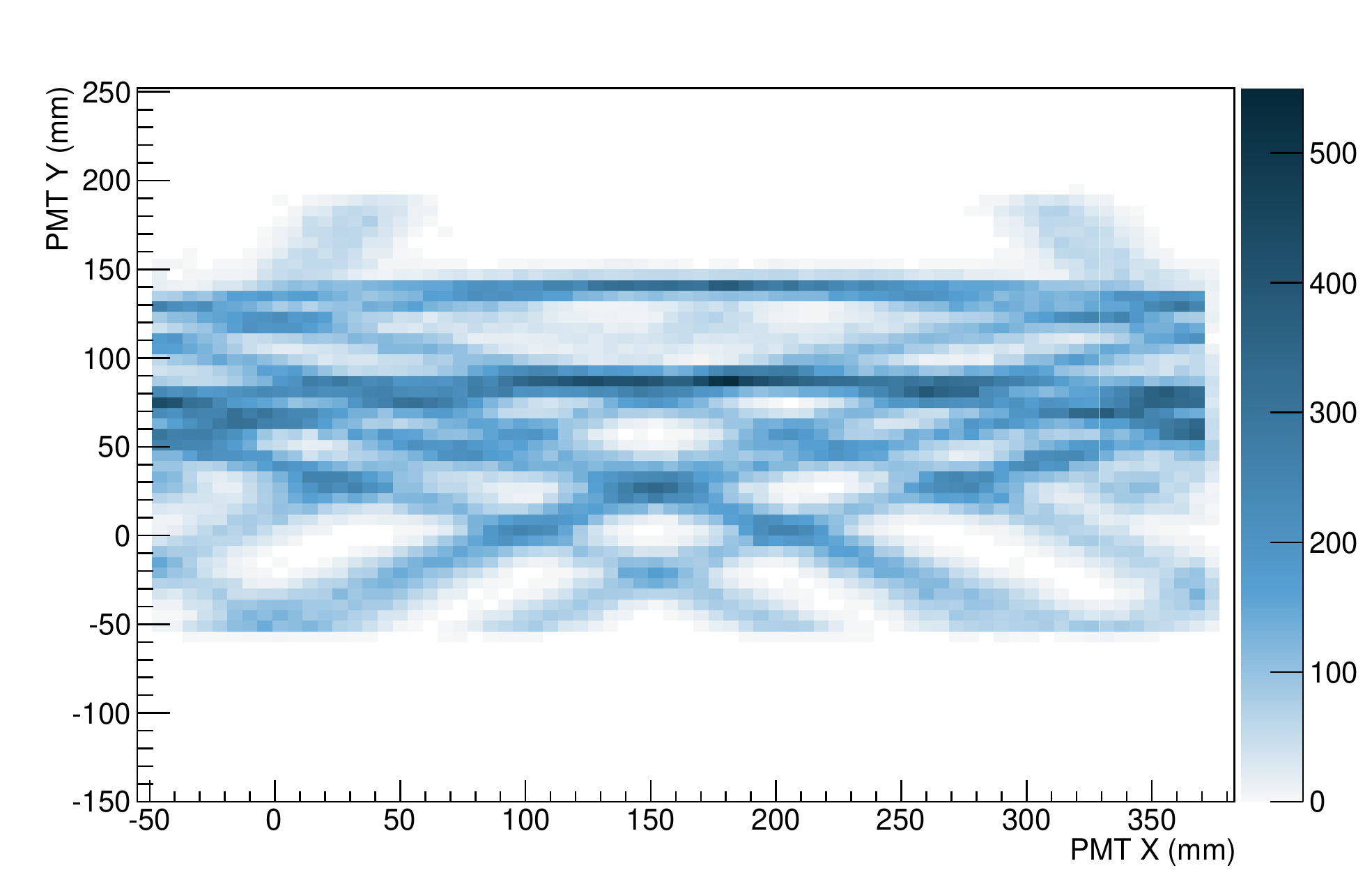}
	\caption{Images obtained on the PMT plane from 10\,000 pions entering the fused-silica bars at a fixed point in space at an angle of $\theta=4^{\circ},\phi=40^{\circ}$ relative to the normal to the surface of the bars.  Images simulated using (left) \geant and (right) FastDIRC.  
} 
	\label{fig:pmt_patterns}
	\end{center}
\end{figure}

Figure~\ref{fig:pmt_patterns} shows the image on the PMT plane obtained by simulating a large number of pions in a full \geant-based Monte Carlo of the SuperB FDIRC~\cite{superbmc}. 
Here we chose to have the pions enter the fused-silica bars at a fixed point in space and at an angle of $\theta=4^{\circ},\phi=40^{\circ}$ relative to the normal to the surface of the bars as this creates a representative non-trivial pattern. This simulation is based on one developed for the PANDA experiment, but uses the SuperB geometry. 
A significant CPU reduction is achieved by applying the photon-detection efficiency immediately after the photon-generation step, {\em i.e.}, prior to propagating the photons through the fused-silica bars.

\subsection{Look-Up-Table Reconstruction}

A common method used in DIRC reconstruction is the so-called look-up-table (LUT) approach.  
This method involves generating a large number of photons at various angles exiting each bar, and then tracking them to the PMT plane.  
For each hit $\vec{x} \equiv (x,y,t)$ recorded in a given PMT pixel, the LUT provides a list of all possible photon-propagation vectors that could have led to a photon hitting the pixel. 
Plotting the cumulative distribution of possible \thetac values for all photons associated to a charged particle -- as obtained from the LUT and the particle's trajectory as measured by the tracking system -- provides a means for measuring \thetac.\footnote{Here we neglected a minor complication.  Typically, each pixel could have been hit by photons emitted from the same bar but at multiple angles. It is common that a LUT algorithm will enter the mean of all possible \thetac values for each pixel into the cumulative distribution to avoid providing more weight to high-occupancy pixels.}
Where each photon exits the bar is dependent upon where its parent particle entered the bar.  Since this information is not known when generating the LUT (which is done prior to processing the data), each photon is assumed to exit from the center of the bar face. 
Therefore, a LUT algorithm produces per-bar images rather than per-particle ones. 

\begin{figure}[]
	\begin{center}
	\includegraphics[width=0.49\textwidth]{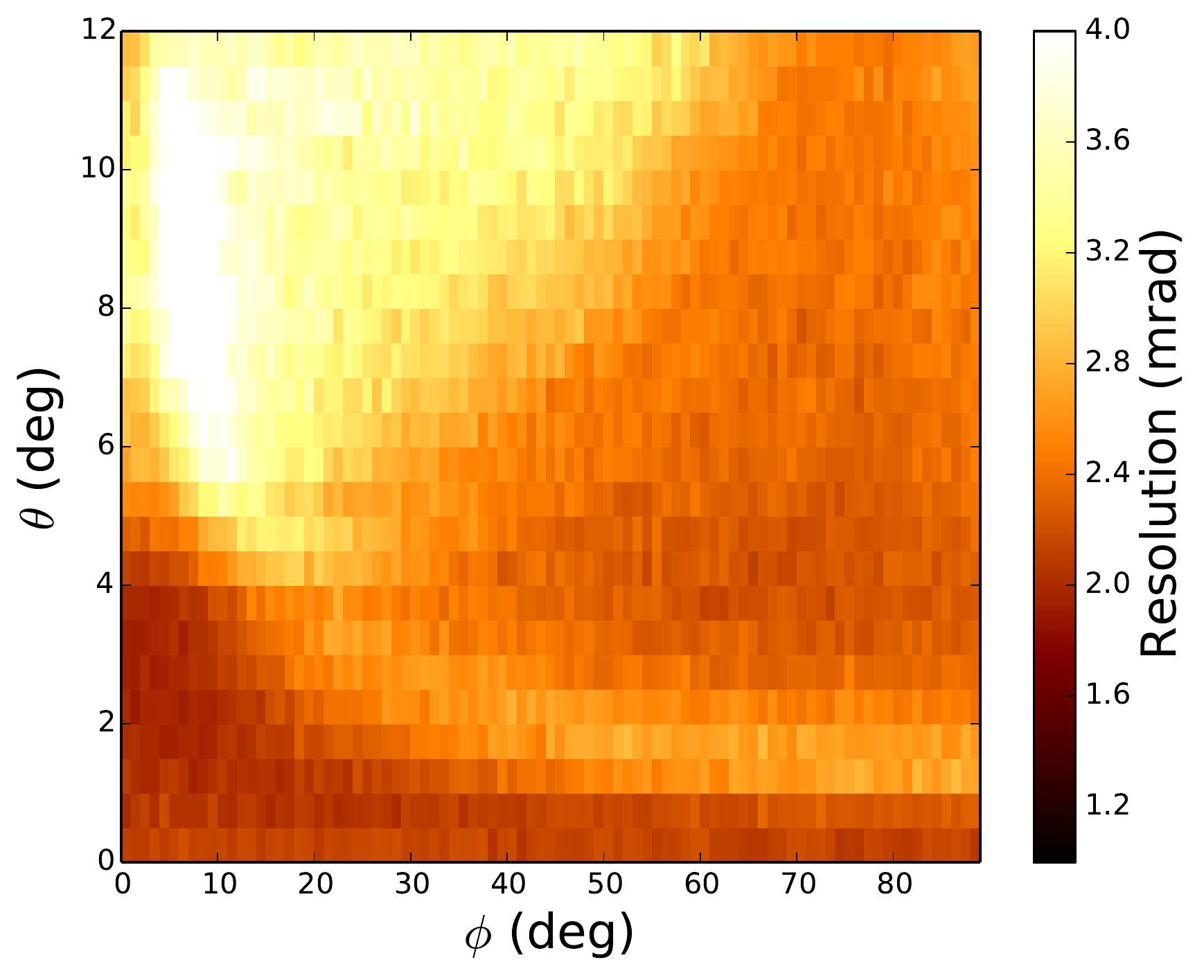}
	\includegraphics[width=0.49\textwidth]{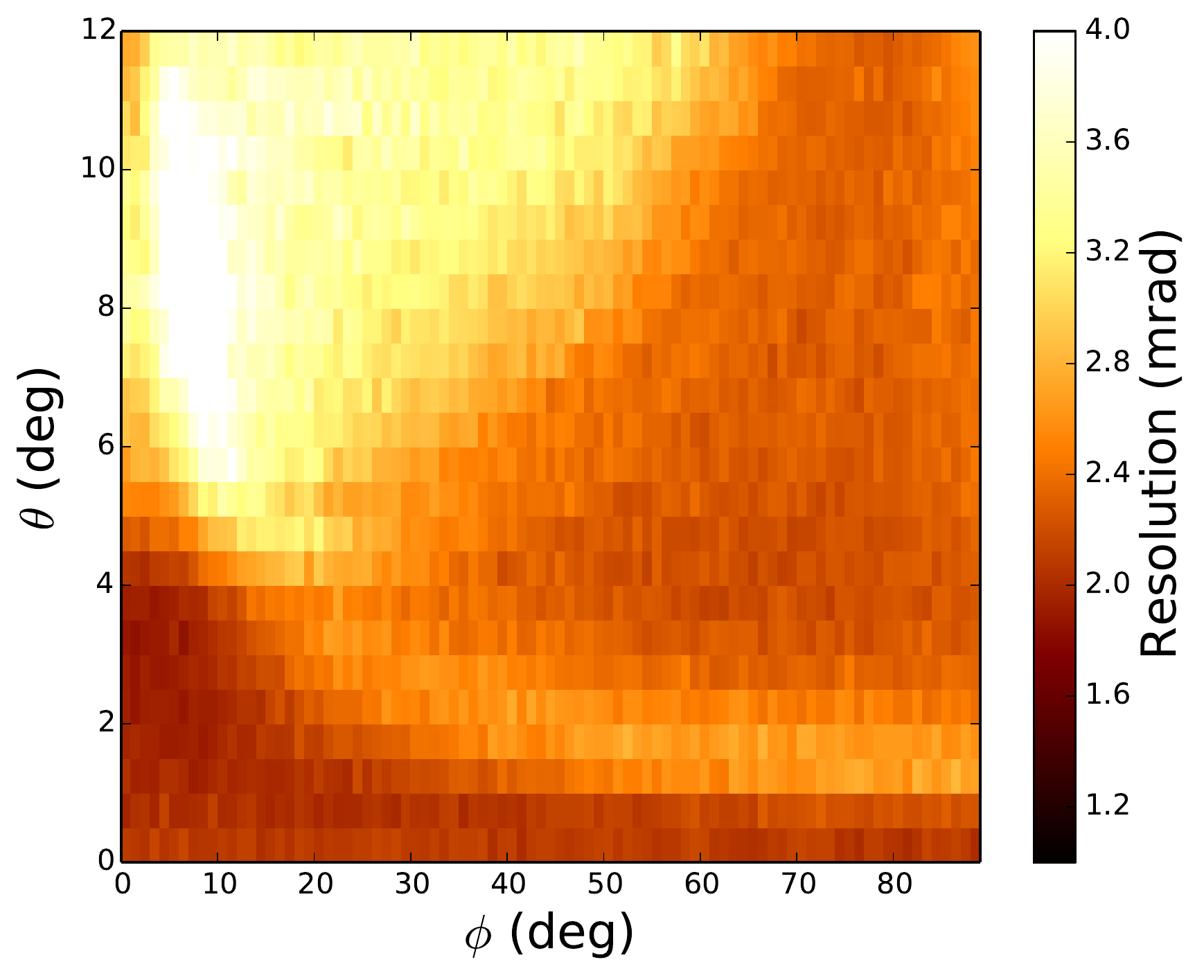}
	\caption{Resolution on \thetac obtained using a LUT algorithm where timing information from the PMTs is (left) only used to separate forward and backward photons and (right) where a so-called chromatic correction is also applied.
        {\em N.b.}, $\theta_C$ resolution is presented per-particle and not per-Cherenkov-photon.
} 
	\label{fig:perf_lut}
	\end{center}
\end{figure}

Figure~\ref{fig:perf_lut} shows the resolution on $\theta_C$ obtained using the LUT algorithm where timing information from the PMTs is: only used to separate forward (emitted towards the PMT plane) and backward (emitted towards the mirrored bar end) photons; and where a so-called chromatic correction is also applied.
The chromatic correction uses the measured time-of-propagation (TOP) of each photon to correct for the fact that Cherenkov photons produced with different wavelengths $\lambda$ are emitted at different angles (see Ref.~\cite{superb} for details). 
Our results obtained using the LUT-based approach for the SuperB prototype design agree with those reported in Ref.~\cite{superb}.

\section{FastDIRC}

The most time-consuming part of the \geant-based DIRC simulation is propagating the Cherenkov photons through the fused-silica bars.  The FastDIRC package uses a novel {\em billiard} approach to this problem, which is described first in this section.  Next, the full FastDIRC simulation algorithm is outlined, followed by description of the reconstruction algorithm.  

\subsection{The Billiard Method}

\begin{figure}[]
	\begin{center}
	\includegraphics[width=9.cm]{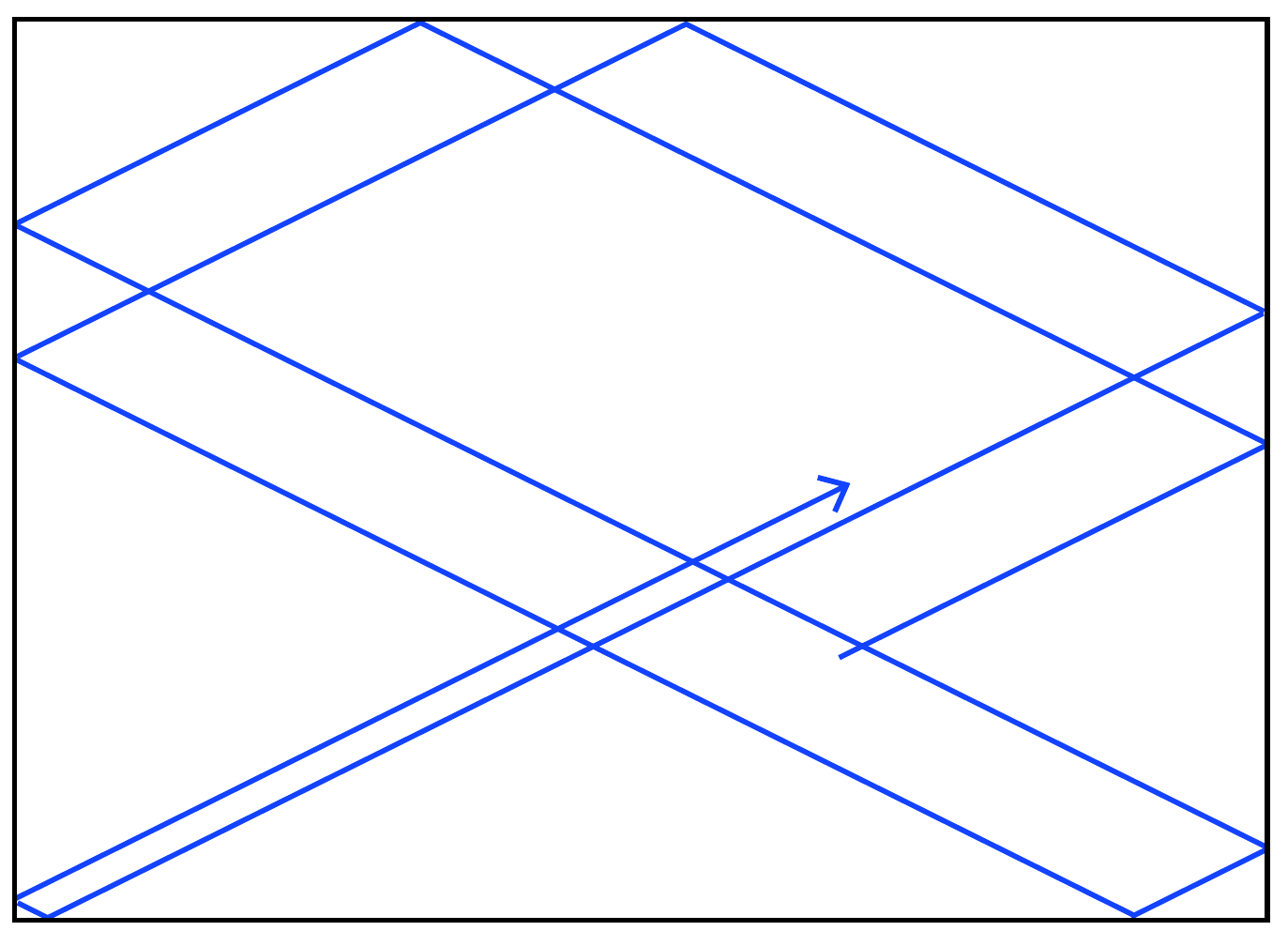}
	\includegraphics[width=9.cm]{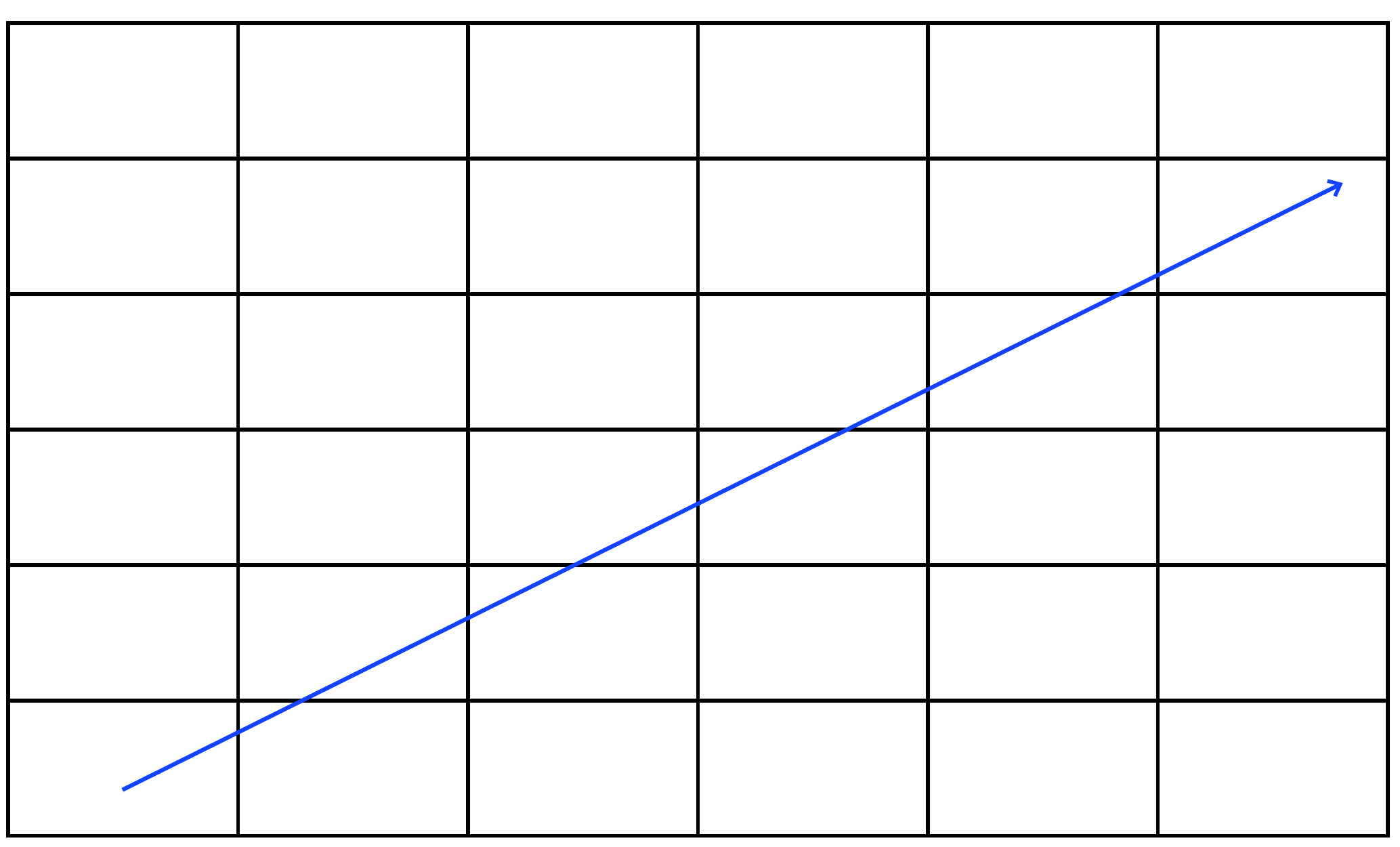}
	\caption{(top) View looking down the long axis of a fused-silica bar as a photon bounces inside it (moving either towards or away from the viewer). (bottom) Same as the top figure but mapped onto a straight-line trajectory through a tiled plane (in this toy example, the photon is reflected 9 times).  
} 
	\label{fig:bounces}
	\end{center}
\end{figure}

Each photon internally reflects $\mathcal{O}(100)$ times as it propagates through a fused-silica bar. Fully simulating each bounce is extremely costly in terms of CPU.  
As an alternative approach, we have developed the billiard method:
\begin{enumerate}
\item The complicated path that each photon takes through a bar is mapped onto a straight-line trajectory through a tiled plane (see Fig.~\ref{fig:bounces}); any rectangular prism can be mapped to a simple tiling of a plane.
\item The distance traveled in the direction defined by the long axis of the bar $\ell_{\parallel}$ is known; depending on which direction the photon is emitted, it is either the distance to the end of the bar or the distance to the mirrored end plus the total bar length. 
\item Given $\ell_{\parallel}$ and the initial (emitted) photon trajectory, the total distance traveled in the orthogonal plane $\ell_{\perp}$ is easily obtained, which permits constructing an unfolded diagrammatic representation of the photon's propagation as in Fig.~\ref{fig:bounces}.
\item The number of reflections is given by the number of intersected cell boundaries, which is obtained via integer division of the path length and cell size.  The final direction of the photon is determined by whether an even or odd number of reflections occurred, while its position exiting the bar is given by the remainder of the integer division. 
\end{enumerate}
Following this simple approach, each Cherenkov photon can be transported directly from its emission point to the end of the bar using minimal CPU power.  

\subsection{Monte Carlo}

The following outline provides an overview of the major steps involved in the FastDIRC simulation for a charged particle with a given momentum $\vec{p}_0$ and spatial entrance point to the bars $\vec{x}_0$:
\begin{enumerate}
\item The particle is propagated through the fused-silica bar starting from $\vec{x}_0$ with direction $\hat{p}_0$.  Multiple scattering of the particle within the bar is simulated by treating the bar as $n_{\rm seg}^{\rm ms}$ segments.  After exiting each segment, the trajectory is augmented using the full Moliere scattering theory, which includes the non-Gaussian Rutherford tails.  The value of $n_{\rm seg}^{\rm ms}$ is configurable, but we find negligible dependence for $n_{\rm seg}^{\rm ms} \gtrsim 10$ for the \babar DIRC bars. 
\item Cherenkov photons are emitted randomly along the particle's path through the bar.  This automatically accounts for the effect on the image due to the thickness of the bars. 
\item The photon-wavelength spectrum generated is the product of the inherent $1/\lambda^2$ Cherenkov-production spectrum and the expected transmission and detection efficiency spectra.  Since the PMT efficiency and in-transport photon-loss distributions are included in the generated spectrum, any photon that reaches an active area of the PMT plane is taken to have perfect detection efficiency.  This approach avoids tracing a large number of photons through the bars and focusing block that end up not being detected.
\item The value of $\theta_C$ for each photon is determined by the particle type and by $\lambda$. 
\item Each photon is analytically traced through the bars using the novel billiard method described in detail above. The rectangular shape of the fused-silica bars is accounted for here.  
\item Upon exiting the bars, the photon direction is smeared using a Gaussian distribution with $\sigma = 3$\,mrad to account for distortion that arises due to bar imperfections as was done at \babar~\cite{babar}.  Characterizing individual bar imperfections has to-date not been achieved; therefore, all DIRC simulations must apply a similar smearing.  In {\sc Geant}, this can be achieved by tuning the surface-finish parameter for the fused-silica bars.  
\item Finally, the photons are analytically traced through the focusing block to the PMT plane and the effects of pixelization and timing resolution are applied to obtain the reconstructed photon information.  This final step is straightforward but also tedious.  It must also account for optical-medium effects if fused-silica is not used.  We have implemented several optical configurations, the details of which can be found on \github.     
\end{enumerate}
Figure~\ref{fig:pmt_patterns} shows that FastDIRC reproduces the PMT images of the full \geant-based Monte Carlo simulation, and it does so in more than $10\,000$ times less CPU time. 

\subsection{Reconstruction}

FastDIRC makes it possible to simulate per-track PDFs for any particle-identity hypothesis that is to be considered.  
This is done by generating -- as described in the previous subsection -- the expected photon-hit distributions assuming the particle is of a given type; however, in this case $n_{\gamma}^{\rm sim} = \mathcal{O}(10\,000) \times n_{\gamma}^{\rm data}$ are produced, {\em i.e.}\ $\mathcal{O}(10\,000)$ times more photons than are expected for a single particle are generated. 
Using the set of $\{\vec{x}_i\}$ values of these photons, a kernel density estimate (KDE) of the PDF for each assumed particle type is obtained as
\begin{equation}
f(\vec{x}) \propto \sum\limits_i^{n_{\gamma}^{\rm sim}} \mathcal{G}(|\vec{x}-\vec{x}_i|),
\end{equation}
where $\mathcal{G}$ is Gaussian function with a mean of zero and width to be optimized for each detector (roughly $1-2$ times the pixel size or timing resolution, but also depends on $n_{\gamma}^{\rm sim}$). 
The choice of a Gaussian function is arbitrary, and many other choices can be found in the literature (see, {\em e.g.}, Ref.~\cite{kde}).
We have tried several options and found minimal dependence on this choice.
  
Using these PDFs and the detected PMT hits associated to a particle, the likelihood is easily obtained for each particle-identity hypothesis. 
Arbitration between particle types is then performed using the likelihood ratio, or equivalently the difference in the log of the likelihoods $\Delta\log{\mathcal{L}}$, formed from the two hypotheses. 
Since this approach naturally includes timing information, the so-called chromatic effect (wavelength dependence of $\theta_C$) is accounted for optimally given the timing resolution. 
In an actual data analysis, $\Delta\log{\mathcal{L}}$ is typically the desired feature to use in event classification; however, in this study, we convert the KDE-based performance into a $\theta_C$ resolution for ease of comparison with the LUT-based results (see Appendix~A for details).  

Figure~\ref{fig:perf_kde} shows the $\theta_C$ resolution obtained using the FastDIRC KDE-based reconstruction algorithm as a function of the charged-particle trajectory relative to the normal to the surface of the bar.  
For all trajectories, the FastDIRC approach achieves a better resolution than the standard look-up-table algorithm. 
This is possible because FastDIRC permits building a unique KDE PDF for each particle, whereas the LUT is a per-bar approximation.  
The mean improvement in the $\theta_C$ resolution integrated over all relative track directions is about 30\%. 

The improvement in the $\theta_C$ resolution does not come without a CPU cost, even in FastDIRC.  The KDE-based approach is about 1000 times slower than using the LUT, though unlike the LUT it is not memory intensive.  
This CPU cost is not an issue when optimizing the design of a DIRC-like detector; however, it likely does not permit using the KDE-based approach on every charged particle in every event recorded by a real-world experiment. 
That said, there are many potential ways of improving the CPU performance of FastDIRC and strategies that could be employed to benefit from the improved resolution provided by FastDIRC; these are discussed in the next section.

\begin{figure}[]
	\begin{center}
	\includegraphics[width=0.49\textwidth]{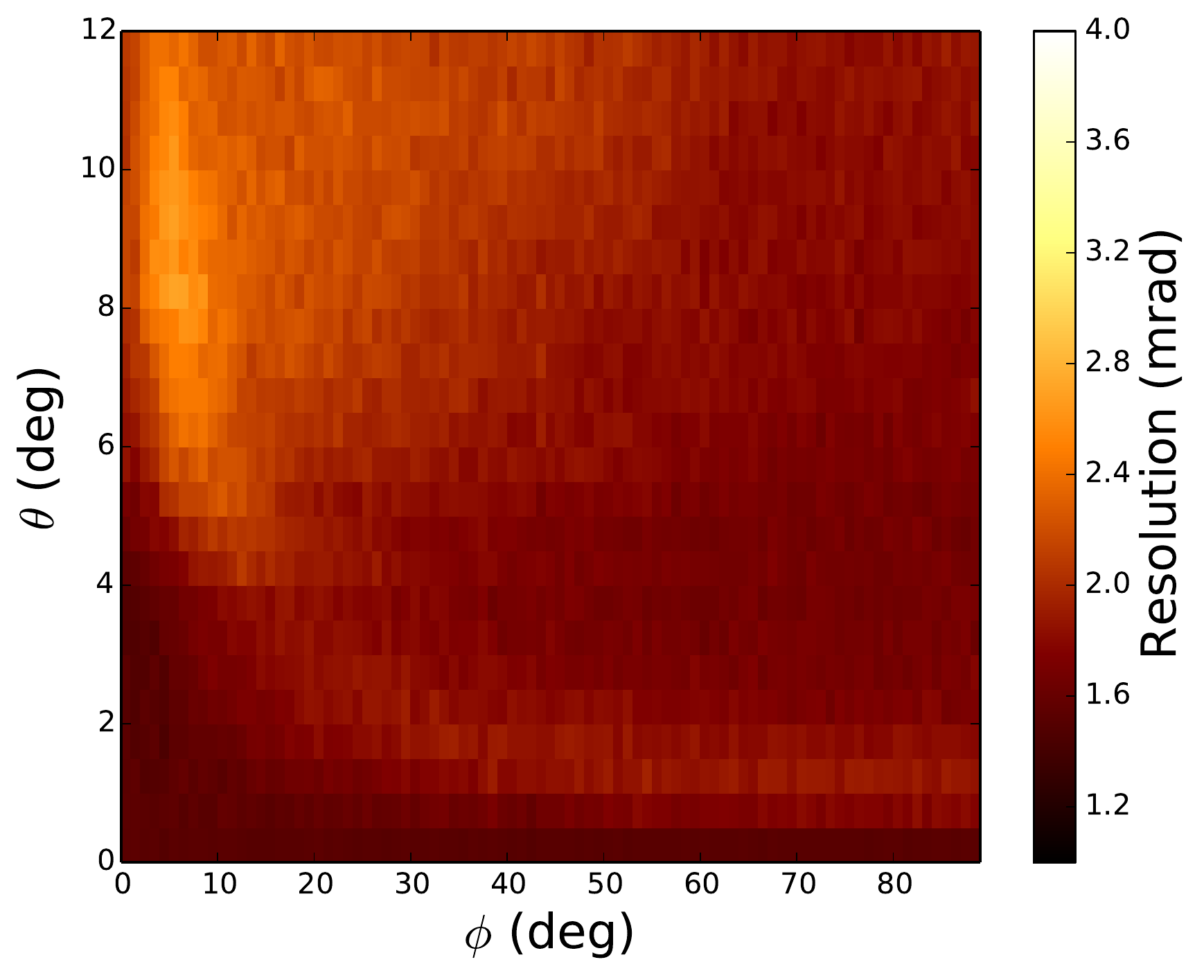}
	\includegraphics[width=0.49\textwidth]{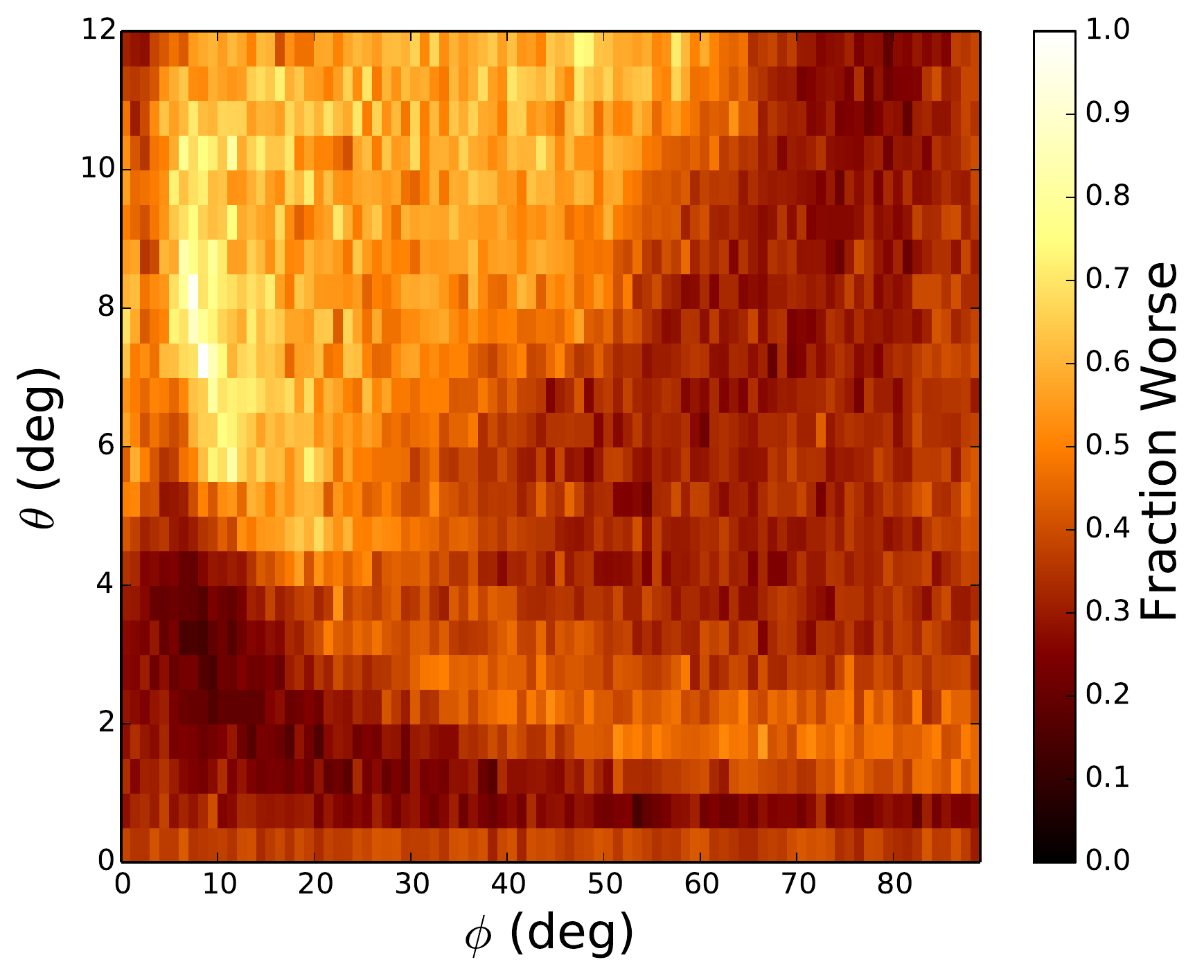}
	\caption{(left) Resolution on $\theta_C$ obtained using the KDE-based method. 
          (right) Relative gain in resolution on $\theta_C$ using the KDE-based versus LUT-based methods, {\em i.e.}, $\left[\sigma_{\rm LUT}(\theta_C)-\sigma_{\rm KDE}(\theta_C)\right]/\sigma_{\rm KDE}(\theta_C)$. 
          The KDE-based method, which is only feasible due to the speed performance provided by the FastDIRC algorithm, achieves better resolution for all trajectories.
} 
	\label{fig:perf_kde}
	\end{center}
\end{figure}

\section{Potential Improvements}

A number of potential improvements could be implemented to increase the speed or usability of the FastDIRC algorithm:
\begin{enumerate}
\item We have implemented custom tracing algorithms for the post-bar optics of \babar, SuperB, and {\sc GlueX}.  Other experiments could be included to expand usability. 
\item Photons are currently randomly sampled but a uniformly stepped sampling would be faster and likely provide equivalent performance. 
\item There are numerous uses of trig and other CPU-intensive functions that could be replaced by approximations as in many cases the argument is restricted to a small domain. 
\item It may be possible to speed up the KDE-based reconstruction by simulating a much smaller number of photons combined with performing an interpolation to build the KDE PDF; however, this is challenging to do without a loss in $\theta_C$ resolution.  The rectangular bar geometry produces many irregular features that can only be captured by simulating a large number of photons.  For example, the so-called kaleidoscope effect does not lend itself well to any interpolation method. 
\item A more viable option for speeding up the KDE-based reconstruction is to generate the KDE PDF using only as many photons as are needed for a given particle-identification arbitration taking into account the uncertainty due to the number of photons used.
\item A LUT-KDE hybrid approach could also be employed, where the LUT is first tried and only followed by the more-precise KDE-based algorithm for cases where the improved precision is required.
\end{enumerate}
Future improvements will be documented in the code at \github.

\section{Summary \& Discussion}

In summary, we presented a novel fast Monte Carlo and reconstruction algorithm for DIRC detectors. 
Our FastDIRC algorithm employs a novel semi-analytic approach to transporting Cherenkov photons to the end of a rectangular fused-silica bar making it more than $10\,000$ times faster than a \geant-based simulation.  
This facilitates using a per-track KDE-based DIRC-reconstruction algorithm that improves the Cherenkov-angle resolution by $\approx 30\%$ compared to existing algorithms. 
The FastDIRC code is available open source at \github. 

\acknowledgments

We thank the {\sc GlueX} collaboration for providing feedback during the development of FastDIRC, and especially C.~Fanelli, B.~Guegan, M.~Patsyuk, and J.~Stevens.  We are grateful to R.~Dzhygadlo and M.~Patsyuk for providing the \geant-based simultion used in this study and to J.~Schwiening for discussions regarding LUT implementations. 
This work was supported by US Department of Energy (DOE) grant DE-SC0010497.

\appendix

\section{$\theta_C$ Resolution from $\Delta \log{\mathcal{L}}$}

\begin{figure}[t]
        \centering
           \includegraphics[height=5.1cm]{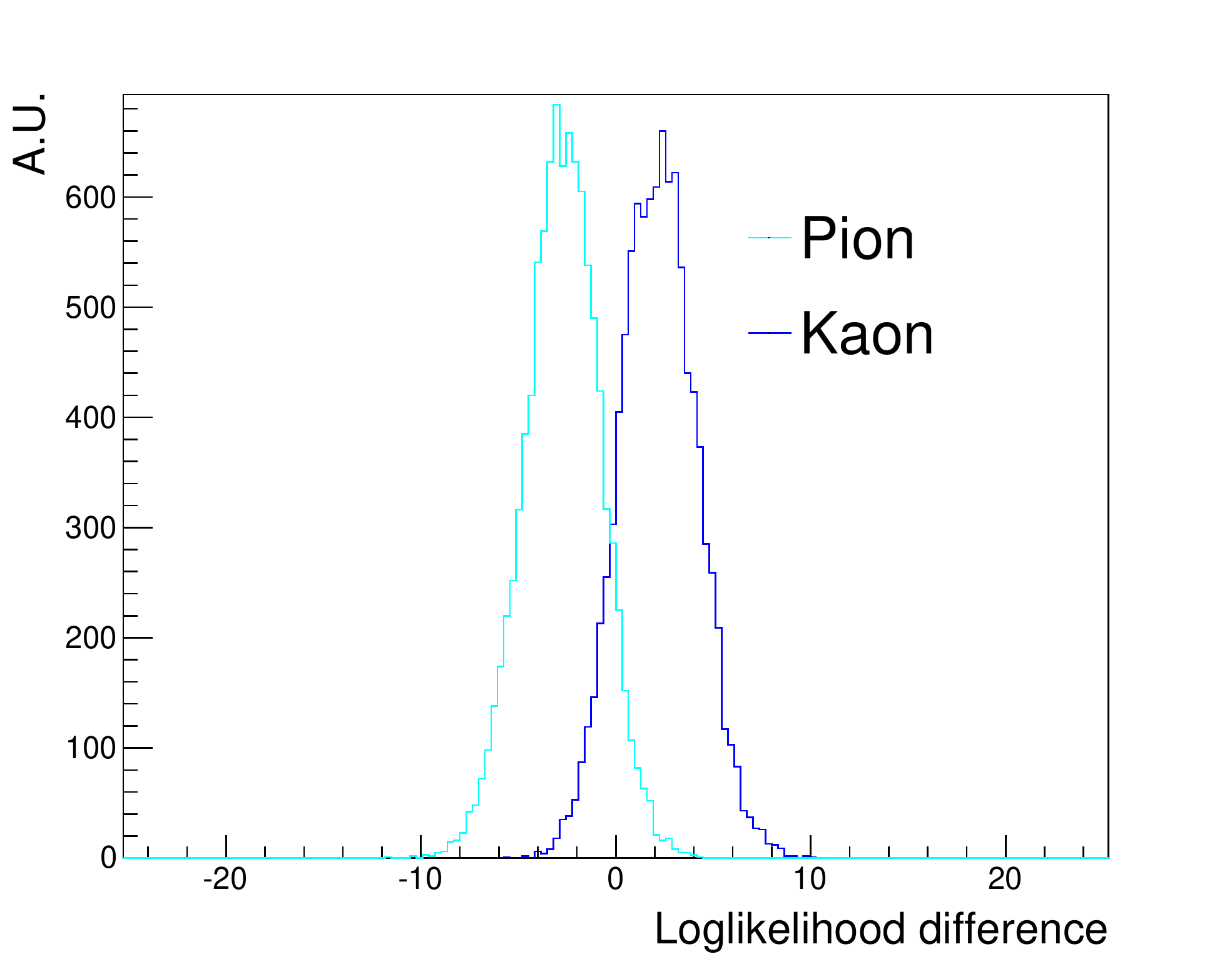}
           \includegraphics[height=5.1cm]{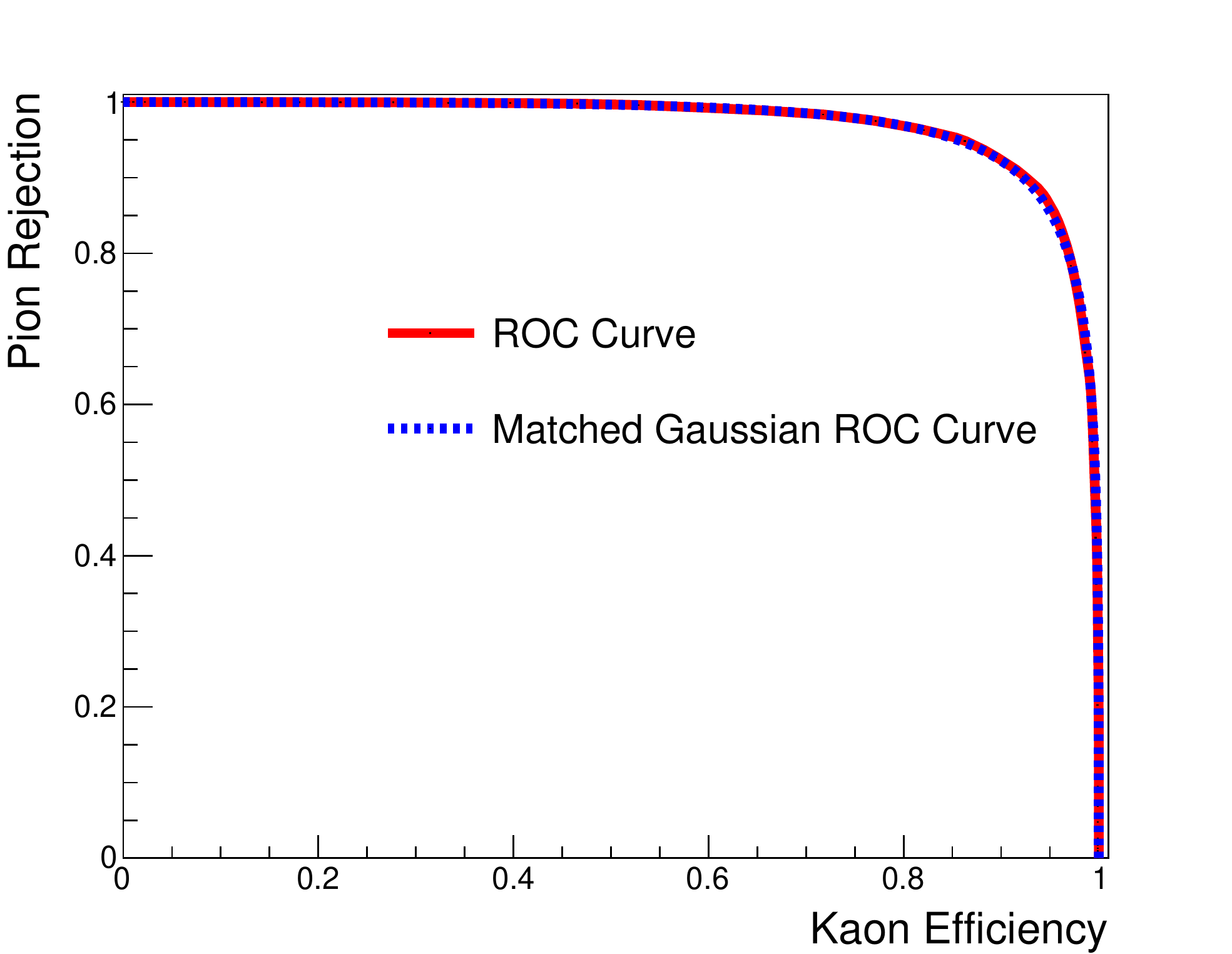}
	\caption{(left) $\Delta\mathcal{L}_{K\pi} \equiv \log{\mathcal{L}_K} - \log{\mathcal{L}_{\pi}}$ for 10\,000 kaons and 10\,000 pions each with $p=5$\,GeV/$c$, $\theta = 4^{\circ}$ and $\phi = 40^{\circ}$.   
        (right) Pion rejection {\em vs} kaon efficiency ROC curve obtained from the same set of kaons and pions. 
} 
	\label{fig:dllkpi_roc}
\end{figure}

In an actual data analysis, $\Delta\log{\mathcal{L}}$ is typically the desired feature to use in event classification; however, in this study, we convert the KDE-based performance into a $\theta_C$ resolution for ease of comparison with LUT-based algorithms.  This is done by considering particles with $p = 5$\,GeV, which is near the edge of the reach of the SuperB FDIRC $K-\pi$ separation.  
Figure~\ref{fig:dllkpi_roc} shows $\Delta\mathcal{L}_{K\pi} \equiv \log{\mathcal{L}_K} - \log{\mathcal{L}_{\pi}}$ for an ensemble of kaons and pions with identical $\vec{p}$.  
The corresponding ROC curve is also shown (ROC curves show background rejection {\em vs} signal efficiency).  
The resolution on $\theta_C$ is extracted by producing the ROC curve for two Gaussian distributions whose means are separated by $\theta_C^K - \theta_C^{\pi}$ (the nominal Cherenkov angles in fused silica for a kaon and pion with this momentum) each with a width $\sigma$.  The value of $\sigma$ is varied until the ROC curve matches the one obtained from the KDE-based $\Delta\mathcal{L}_{K\pi}$ method (technically, we find the value of $\sigma$ that produces the same integral under the ROC curve).  
 The best match from an ideal Gaussian ROC curve is also shown in Fig.~\ref{fig:dllkpi_roc}, which demonstrates that in the absence of tracking uncertainties, the DIRC performance is nearly Gaussian. 

\begin{thebibliography}{99}
\bibitem{babar} I. Adam {\it et al.},  Nucl. Instr. and Meth. {\bf A538}, 281 (2005).
\bibitem{superb}  J. Va'vra {\it et al.},  Nucl. Instr. and Meth. {\bf A718}, 541 (2013); B. Dey {\em et al.}, Nucl. Instr. and Meth. {\bf A775}, 112 (2015).
\bibitem{belle2} 
 K.~Inami {\it et al.} [Belle-II PID Group collaboration],
  Nucl.\ Instrum.\ Meth.\ A {\bf 766}, 5 (2014).
\bibitem{gluex} 
J.~Stevens {\it et al.},
JINST {\bf 11}, C07010 (2016). 
\bibitem{torch}
 R.~Forty and M.~Charles,
  LHCb-PUB-2009-030, CERN-LHCb-PUB-2009-030.
\bibitem{panda} 
J. Schwiening, 
Nucl. Instrum. Meth. A {\bf 639} 315 (2011);
R. Dzhygadlo {\em et al.}, JINST {\bf 11}, C05013 (2016).
\bibitem{geant}
     S.~Agostinelli {\it et al.} [Geant4 collaboration],
Nucl.\ Instrum.\ Meth.\ A {\bf 506} 250 (2003).
\bibitem{superbmc} The \geant-based SuperB FDIRC Monte Carlo geometry was implemented by R.~Dzhygadlo and M.~Patsyuk. See 
{\tt http://web-docs.gsi.de/\textasciitilde rdzhigad/www/}
\bibitem{kde} 
 K.~S.~Cranmer,
  Comput.\ Phys.\ Commun.\  {\bf 136}, 198 (2001)
  [hep-ex/0011057].
\end{thebibliography}
\end{document}